\documentclass[reprint, two column, prl, aps, superscriptaddress, showkeys]{revtex4-1}
\usepackage{amssymb}
\usepackage{amsmath}
\usepackage{graphicx}
\usepackage{dcolumn}
\usepackage{bm}
\usepackage{color}
\usepackage{ulem}
\usepackage{float}
\usepackage{textcomp}
\restylefloat{table}
\usepackage{verbatim}
\usepackage{epstopdf}
\usepackage[mediumspace,mediumqspace,squaren]{SIunits}

\begin{document}

\title{Geometric Resonance of Four-Flux Composite Fermions }
\date{\today}

\author{Md.\ Shafayat Hossain}
\author{Meng K.\ Ma}
\author{M. A.\ Mueed}
\author{D.\ Kamburov}
\author{L. N.\ Pfeiffer} 
\author{K. W.\ West}
\author{K. W.\ Baldwin}
\affiliation{Department of Electrical Engineering, Princeton University, Princeton, New Jersey 08544, USA}
\author{R.\ Winkler}
\affiliation{Department of Physics, Northern Illinois University, DeKalb, Illinois 60115, USA}

\author{M.\ Shayegan}
\affiliation{Department of Electrical Engineering, Princeton University, Princeton, New Jersey 08544, USA}

\begin{abstract}
Two-dimensional interacting electrons exposed to strong perpendicular magnetic fields generate emergent, exotic quasiparticles phenomenologically distinct from electrons. Specifically, electrons bind with an even number of flux quanta, and transform into composite fermions (CFs). Besides providing an intuitive explanation for the fractional quantum Hall states, CFs also possess Fermi-liquid-like properties, including a well-defined Fermi sea, at and near even-denominator Landau level filling factors such as $\nu=1/2$ or $1/4$. Here, we directly probe the Fermi sea of the rarely studied four-flux CFs near $\nu=1/4$ via geometric resonance experiments. The data reveal some unique characteristics. Unlike in the case of two-flux CFs, the magnetic field positions of the geometric resonance resistance minima for $\nu<1/4$ and $\nu>1/4$ are symmetric with respect to the position of $\nu=1/4$. However, when an in-plane magnetic field is applied, the minima positions become asymmetric, implying a mysterious asymmetry in the CF Fermi sea anisotropy for $\nu<1/4$ and $\nu>1/4$. This asymmetry, which is in stark contrast to the two-flux CFs, suggests that the four-flux CFs on the two sides of $\nu=1/4$ have very different effective masses, possibly because of the proximity of the Wigner crystal formation at small $\nu$.
\end{abstract} 

\maketitle

Ultra-low-disorder two-dimensional electron systems (2DESs) subjected to a perpendicular magnetic field ($B_{\perp}$) give rise to a plethora of quantum many-body phases of matter. Many of these phases can be understood based on composite fermions, quasiparticles comprised of an electron and an even number of flux quanta \cite{Jain.PRL.1989, Jain.2007, Halperin.PRB.1993}. Near Landau level (LL) filling factor $\nu=1/2$, e.g., an electron merges with two flux quanta to form a two-flux composite fermion ($^2$CF). While the electron system is highly interacting and is in a high $B_{\perp}$, the $^2$CFs behave as essentially non-interacting particles and only feel an \textit{effective} magnetic field $B^{*}=B-B_{\nu=1/2}$, where $B_{\nu=1/2}$ is the field at  $\nu=1/2$.  Importantly, these $^2$CFs occupy a Fermi sea at $\nu=1/2$ and can execute cyclotron motion near $\nu=1/2$ at small $B^{*}$, similar to their fermion counterparts near $B=0$ \cite{Halperin.PRB.1993}.  With the application of a one-dimensional periodic perturbation to the 2DES, if the $^2$CFs can complete a cyclotron orbit ballistically, then they exhibit a geometric resonance (GR) when their orbit diameter equals the period of the perturbation. Such a resonance provides a direct and quantitative way to explore some of the fundamental properties of $^2$CFs \cite{Willett.PRL.1993, Kang.PRL.1993, Smet.PRL.1999, Kamburov.PRL.2013, Kamburov.PRB.2014, Kamburov.PRL.2014}. For example, recent GR measurements of $^2$CF Fermi sea revealed an unexpected asymmetry between the two sides of $\nu=1/2$ \cite {Kamburov.PRL.2014}. This asymmetry, and more generally the question of particle-hole symmetry, inspired renewed interest in the physics of a half-filled LL \cite {Barkeshli.PRB.2015, Kachru.PRB.2015, Son.PRX.2015, Balram.PRL.2015, Balram.PRB.2016, Wang2.PRB.2016, Wang3.PRB.2016, Mulligan.PRB.2016, Geraedts.Science.2016, Zucker.PRL.2016, Balram.PRB.2017, Wang.PRX.2017, Cheung.PRB.2017, Pan.NatPhys.2017, Geraedts.PRL.2018, Goldman.PRB.2018, Son.Annul.Rev.Cond.Mat.Phys.2018, Mitra.preprint}. Notable among the new studies is the theory involving a Dirac fermion description \cite {Son.PRX.2015, Wang2.PRB.2016, Wang3.PRB.2016, Mulligan.PRB.2016, Geraedts.Science.2016, Zucker.PRL.2016, Wang.PRX.2017, Cheung.PRB.2017, Pan.NatPhys.2017, Geraedts.PRL.2018, Goldman.PRB.2018, Son.Annul.Rev.Cond.Mat.Phys.2018, Mitra.preprint}.

Qualitatively similar to the case of $\nu=1/2$, at $\nu=1/4$ electrons merge with four flux quanta and form a four-flux CF ($^4$CF) Fermi sea. Unlike $\nu=1/2$, there is no obvious particle-hole symmetry at $\nu=1/4$ \cite{footnote}. This provides motivation for studies of $^4$CFs whose physics could be distinct from $^2$CFs. However, measurements of $^4$CFs are very scarce \cite{Leadley.PRL.1994, Yeh.PRL.1999, Pan.PRB.2000}, partly because they require very high magnetic fields, and also because of the proximity of $\nu=1/4$ to the Wigner crystal formation near $\nu=1/5$ \cite{Andrei.PRL.1988, Jiang.PRL.1990, Goldman.PRL.1990}. Therefore, many fundamental questions have remained unanswered: Do $^4$CFs have properties similar to the $^2$CFs? Do $^4$CFs show an asymmetry in the field positions of the GR minima similar to $^2$CFs \cite{Kamburov.PRL.2014}? What happens to the $^4$CF Fermi sea when the Fermi sea for zero-field electrons is highly anisotropic? Our GR measurements reported here provide answers to these fundamental questions, and reveal surprises for $^4$CFs.

\begin{figure}[t!]
\includegraphics[width=0.49\textwidth]{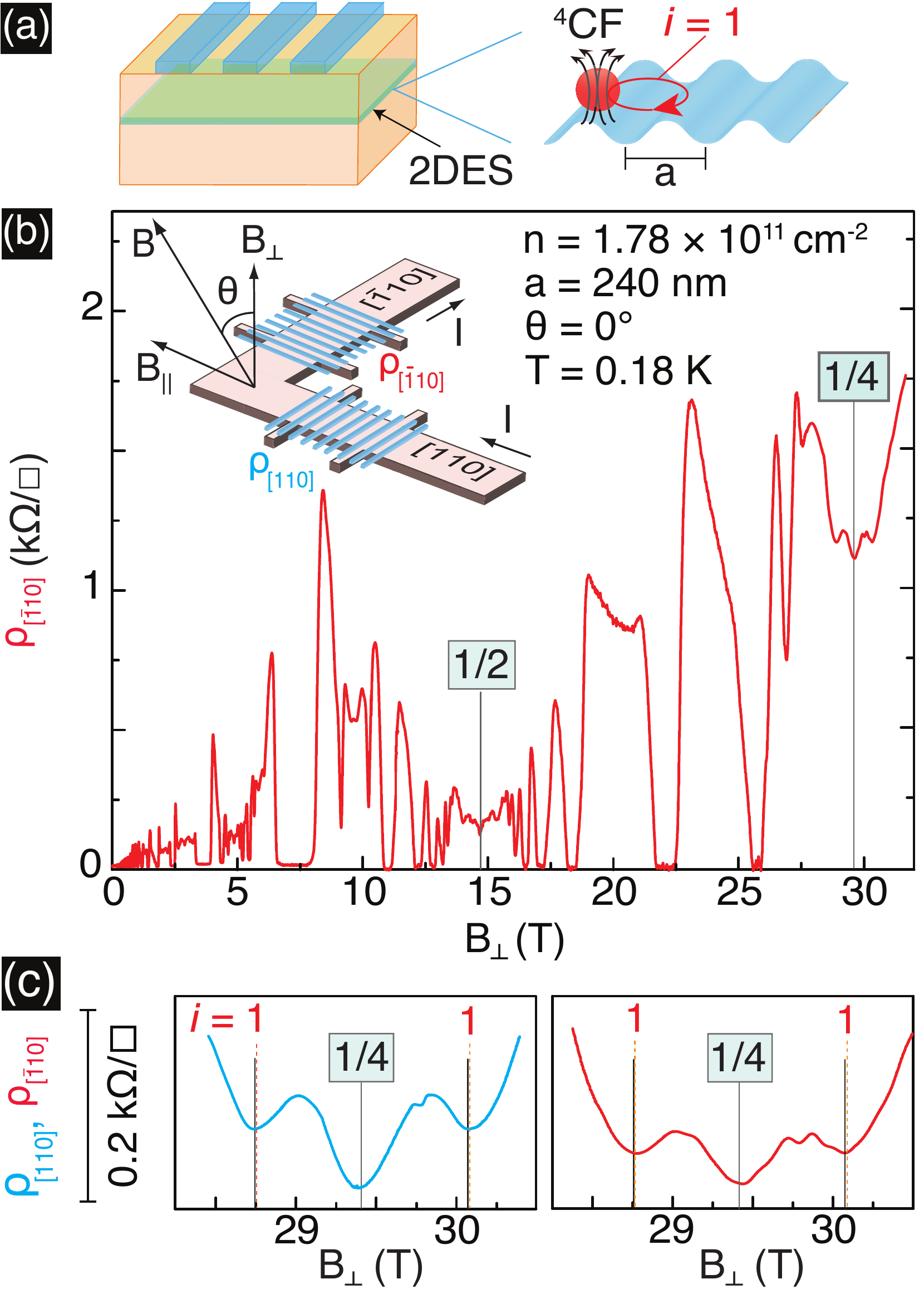}
\caption{\label{fig:Fig1} GR features for $^4$CFs near $\nu=1/4$. (a) Lateral surface superlattice of period $a$, inducing a periodic density perturbation in the 2DES. When the $^4$CFs' cyclotron orbit becomes commensurate with the period of the perturbation, the $i=1$ GR occurs. (b) Magneto-resistance trace revealing GR features near $\nu=1/4$ and $\nu=1/2$. Inset: The L-shaped Hall bar along $[110]$ and $[\bar{1}10]$ directions used for the measurements. (c) Magneto-resistance near $\nu=1/4$ demonstrating the $i = 1$ $^4$CF GR features, resistance minima flanking $\nu=1/4$. Black solid and orange dashed lines mark the $expected$ positions for the $i = 1$ GR for fully spin-polarized $^4$CFs with circular Fermi contour assuming $k_F^*=\sqrt{4\pi n}$ and $k_F^*=\sqrt{4\pi n} \times\sqrt{B/B_{\nu=1/4}}$, respectively. The extra minimum near $B_{\perp}=29.75$ T stems from the $i=2$ GR.}
\end{figure}  

Our experimental platform is a molecular beam epitaxy grown 2DES, with density $n= 1.78\times10^{11}$ cm$^{-2}$ and low-temperature mobility $1.4\times 10^{7}$ cm$^2/$Vs, confined to a modulation-doped, 40-nm-wide, GaAs quantum well \cite{SM}. In our GR measurements, we impose a minute periodic density modulation, the estimated magnitude of which is about $0.5\%$ \cite{Mueedstripe.PRL.2016}. As illustrated in Fig. 1(a), this is achieved by fabricating a one-dimensional superlattice of period $a=240$ nm, consisting of stripes of negative electron-beam resist on the surface of a lithographically-defined Hall bar \cite{ Kamburov.PRB.2014, Skuras.APL.1997, SM, Endo.PRB.2000, Endo.PRB.2001, Endo.PRB.2005, Mueedstripe.PRL.2016, Davies.PRB.1994, Kamburov3/2.PRB.2014, Kamburov.PRL.2014, Mueed.PRB.2017, Mueed.PRL.2015, Mueed.PRL.2016, Shafayat5/2.PRL.2018}. Thanks to the piezoelectric effect in GaAs, the strain from this surface superlattice propagates to the 2DES which is 235 nm underneath the sample surface and leads to a small density modulation. 

\begin{figure}[t!]
\includegraphics[width=.49\textwidth]{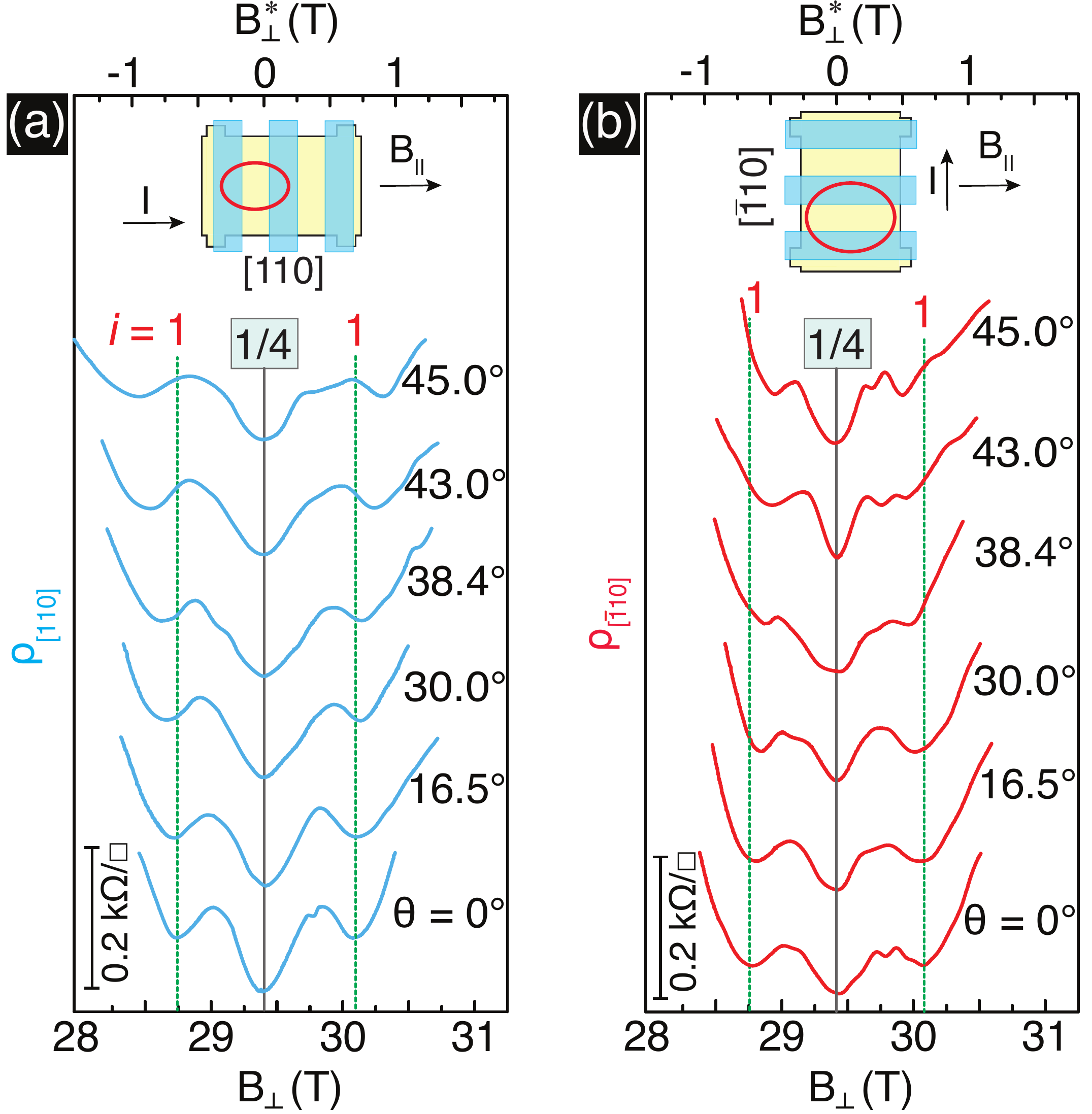}
\caption{\label{fig:Fig2} Tilt evolution of the $^4$CF GR features near $\nu = 1/4$ along (a) $[110]$ and (b) $[\bar{1}10]$ directions. The insets show the orientation of the Hall bars, and the $^4$CF cyclotron orbit for the $i=1$ GR. Magneto-resistance traces are vertically offset for clarity; the tilt angle $\theta$ is given for each trace. The $expected$ positions for the $i=1$ $^4$CF GRs are marked with vertical dotted lines assuming that $k_F^* = \sqrt{4\pi n}$. In both panels, the scale for the applied external field $B_{\perp}$ is shown on bottom while the top scale is the effective magnetic field $B^*_{\perp}$ experienced by the $^4$CFs.}
\end{figure}

Weakly-interacting CFs subjected to an effective perpendicular magnetic field $B^{*}$ execute circular cyclotron motion with an orbit radius of $R_{c}^{*}=\hbar k_F^{*}/eB^{*}$, the size of which is determined by the magnitude of the Fermi wave vector of the CFs, $k_F^{*}$ \cite{footnote1}. If the CFs have a sufficiently long mean-free-path so they can complete a ballistic cyclotron orbit, then a GR occurs when the orbit diameter becomes commensurate with the period ($a$) of the modulation; see Fig. 1(a) for a schematic illustration. More quantitatively \cite{Smet.PRL.1999, Kamburov.PRL.2013, Kamburov.PRB.2014, Kamburov3/2.PRB.2014, Kamburov.PRL.2014, Mueed.PRB.2017}, when $2R_{c}^{*}/a=i+1/4$ ($i=1,2,3,...$), GRs manifest as minima in magneto-resistance at $B_{i}^{*}=2\hbar k_F^{*}/ea(i+1/4)$. Thus, $k_F^{*}$ can be deduced directly from the positions of $B_{i}^{*}$. Such direct measurement of $k_F^{*}$ not only provides a proof for the existence of a CF Fermi sea and a measure of its spin polarization but also enables one to quantitatively investigate how the anisotropy of the electron Fermi sea transfers to the CF Fermi sea \cite{Smet.PRL.1999, Kamburov.PRL.2013, Kamburov.PRB.2014, Kamburov3/2.PRB.2014, Kamburov.PRL.2014, Mueed.PRB.2017, Shafayat5/2.PRL.2018, Mueed.PRL.2015, Mueed.PRL.2016}. Here, we apply this technique in very high magnetic fields (using a 45 T hybrid magnet) to investigate the $^4$CFs near $\nu=1/4$.

We first show in Fig. 1(b) a representative magneto-resistance trace, exhibiting well-developed GR features flanking symmetrically a deep V-shaped minimum at $\nu=1/4$. Figure 1(c) zooms in around $\nu=1/4$. From the period of the modulation, $a=240$ nm, we determine the expected positions for the primary $i = 1$ GR resistance minima according to $B_{i=1}^{*}=2\hbar k_F^{*}/ea(1+1/4)$ where $B_{i=1}^*=B_{i=1}-B_{\nu=1/4}$. We assume a fully spin-polarized CF sea and mark the expected positions for $B_{i=1}$ in Fig. 1(c) considering two possibilities: (i) black solid lines for $k_F^*=\sqrt{4\pi n}$, and (ii) orange dashed lines for $k_F^*$ changing according to the magnetic length, i.e., $k_F^*=\sqrt{4\pi n}\times\sqrt{B/B_{\nu=1/4}}$ \cite{Halperin.PRB.1993, Son.PRX.2015, Wang2.PRB.2016, Wang3.PRB.2016, Wang.PRX.2017, Cheung.PRB.2017, Son.Annul.Rev.Cond.Mat.Phys.2018, Mitra.preprint}. The difference between the expected $B_{i=1}$ for the two assumptions is very small and cannot be resolved in our experiments. From Fig. 1(c), it is clear that the observed GR minima positions are in excellent agreement with the expected $B^*_{i=1}$, confirming that the $^4$CFs near $\nu=1/4$ are fully spin polarized \cite{footnote.n1}. More importantly, unlike the $^2$CF GRs flanking $\nu=1/2$ \cite{Kamburov.PRL.2014}, the GR features for $^4$CFs are quite symmetric around $\nu=1/4$. This is reasonable, considering that the \textit{minority} carrier density, which was found experimentally in Ref. \cite{Kamburov.PRL.2014} to determine $k_F^*$ for $^2$CFs, is the same on the two sides of $\nu=1/4$ and is equal to $n$ \cite{footnote.n2}.

A fundamental question regarding emergent quasiparticles such as CFs in high magnetic fields is how an anisotropy in the Fermi sea of the electrons at zero field affects the CF Fermi sea \cite{Kamburov.PRL.2013, Kamburov.PRB.2014, Gokmen.Natphy.2010, Haldane.PRL.2011, Yang.PRB.2012, Wang.PRB.2012, Yang.PRB.2013, Balram2.PRB.2016, Jo.PRL.2017, Ippoliti1.PRB.2017, Ippoliti2.PRB.2017, Ippoliti3.PRB.2017, Lee.Preprint.2018}. To address this question, we apply an in-plane magnetic field ($B_{||}$) which, through its coupling to the out-of-plane motion of the electrons in a quasi-2D system, severely distorts the Fermi sea of the low-field electrons \cite{Kamburov.PRB.2012, Kamburov.PRB.2013, Mueed.PRL.Fermi2015}. The application of $B_{||}$ shrinks the real-space cyclotron orbit diameter in the in-plane direction perpendicular to $B_{||}$, thereby shrinking the Fermi sea in the direction of $B_{||}$.

The subsequent anisotropy of the CF cyclotron orbit can be determined in a straightforward manner via measuring the positions of the CF GR minima along the two perpendicular arms of the L-shaped Hall bar [inset of Fig. 1(b)]. Since the reciprocal-space ($k$-space) orbits are expected to be a scaled version of the real-space trajectories, rotated by $90^o$  \cite{AM.BOOK}, our GR measurements then directly probe the Fermi sea shape. In our experiments, we tilt the sample so that $B_{||}$ is always along $[110]$, with $\theta$ denoting the angle between the field direction and the normal to the 2D plane [Fig. 1(b) inset].

As seen in Fig. 2, the application of $B_{||}$ affects the positions of the $^4$CF GR minima.  Traces for the two arms of the Hall bar along $[110]$ and $[\bar{1}10]$ are shown in Figs. 2(a) and 2(b). In both panels, the vertical dotted lines mark the expected positions of the $i = 1$ CF GR minima for spin-polarized $^4$CFs with a circular Fermi sea, i.e., $B_{i=1}^{*} = 2\hbar \sqrt{4\pi n}/ ea(1+1/4)$. These lines match the observed positions of the resistance minima for the bottom traces of Fig. 2, which were taken at $\theta=0^o$. When we increase $\theta$ and thereby  $B_{||}$, for the [110] arm [Fig. 2(a)], the positions of the two GR minima shift away from $\nu=1/4$ to larger values of $|B_{\perp}^*|$. In contrast, the GR minima for the $[\bar{1}10]$ arm [Fig. 2(b)] move towards smaller $|B_{\perp}^*|$. Using the field positions of the GR minima along the $[110]$ and $[\bar{1}10]$ directions, we directly extract the magnitude of the Fermi wavevector $\bm{k}_F^*$ along $[\bar{1}10]$ and $[110]$, respectively; we use the expression: $k_F^{*}=B_{i=1}^{*}ea(1+1/4)/2\hbar$.

\begin{figure}[t!]
\includegraphics[width=.49\textwidth]{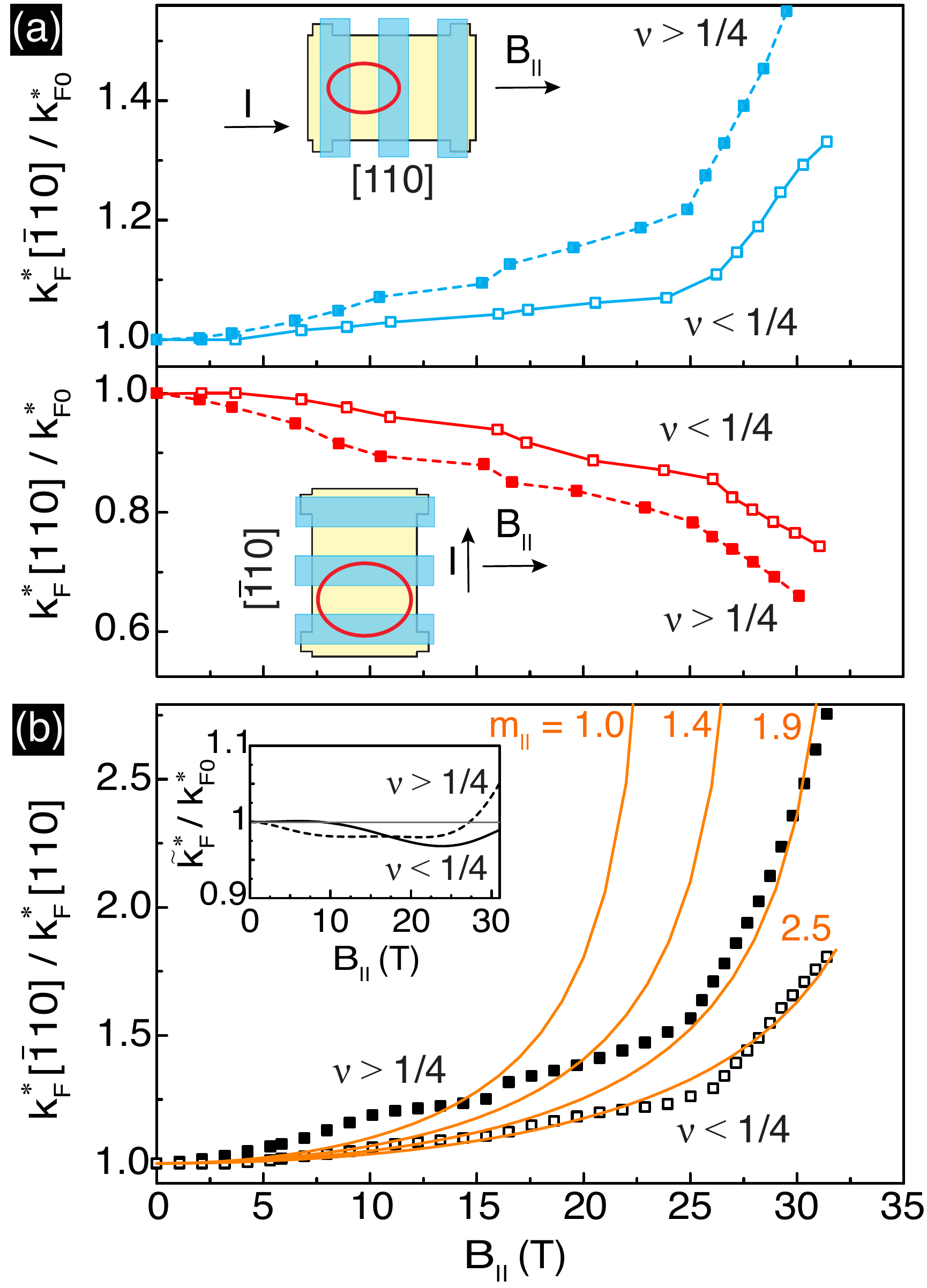}
\caption{\label{fig:Fig3} (a) Normalized $^4$CF Fermi wave vectors $k^*_F$ from the positions of B$^*_{\perp}$ for the primary $^4$CF GR minima along the $[110]$ and $[\bar{1}10]$ directions. Open and filled symbols represent the data for $\nu<1/4$ and $\nu>1/4$, respectively. The typical error bar for the data points is of the order of $3\%$. (b) Anisotropy of the $^4$CF Fermi sea for $\nu<1/4$ (open symbols) and $\nu>1/4$ (filled symbols) deduced from dividing the (interpolated) measured values of $k^*_F$ along $[\bar{1}10]$ by those along $[110]$. Orange lines correspond to the theoretical estimate of the anisotropy using Eq. (1) assuming $m_{||} = 2.5, 1.9, 1.4$, and $1.0$ (see text). Inset: Geometric mean of the measured values of $k^*_F$ ($\tilde{k}^*_F = \sqrt{k^*_F[110]\times k^*_F [\bar{1}10]}$) along the two directions normalized to $k^*_{F0}$ for $\nu<1/4$ and $\nu>1/4$ denoted by solid and dashed lines, respectively. Up to the highest $B_{||}$, $\tilde{k}^*_F/k_F^*\simeq1$ to within $5\%$, implying that the measured Fermi seas are nearly elliptical. }
\end{figure}

\begin{figure}[t!]
\includegraphics[width=.49\textwidth]{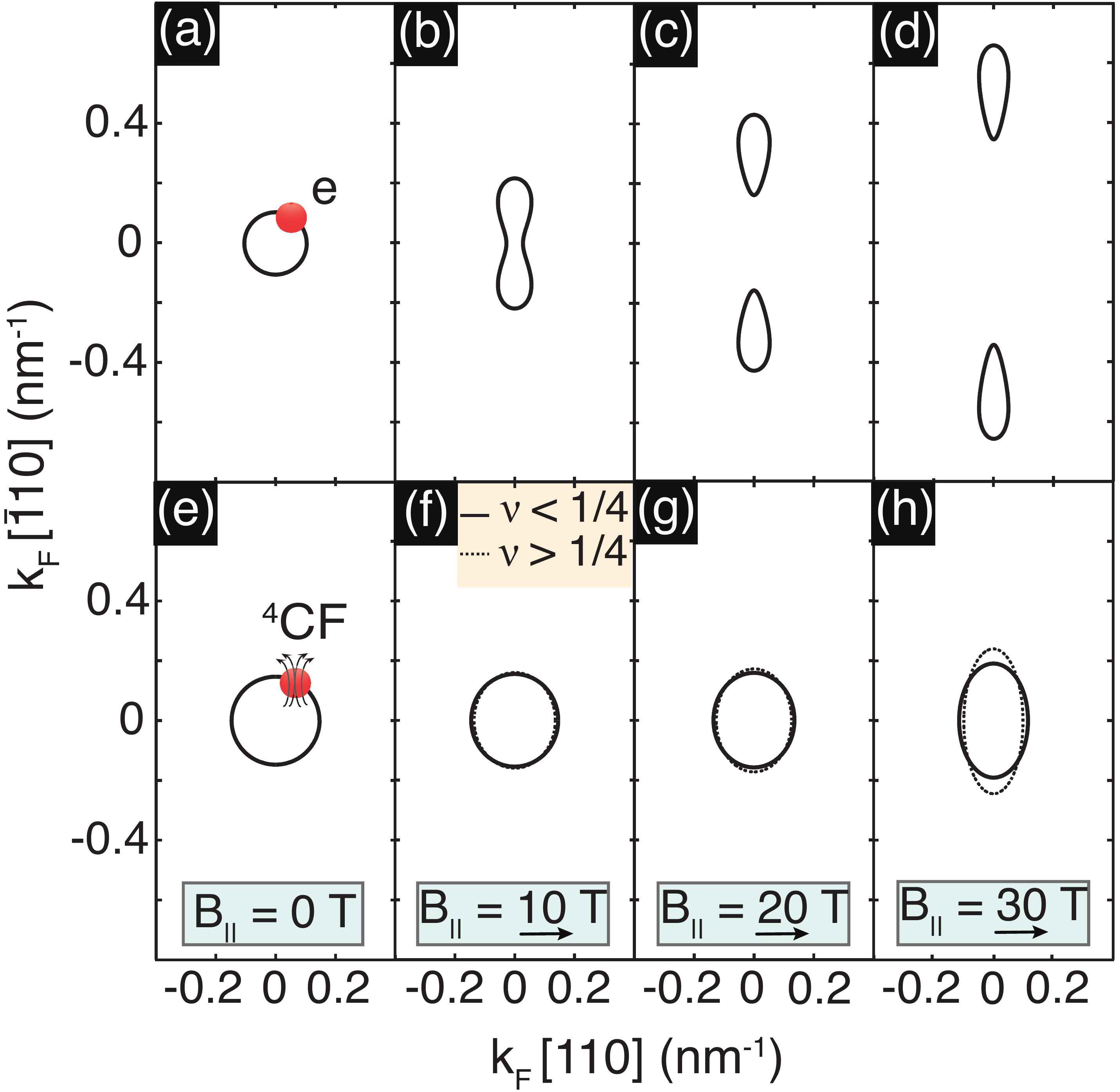}
\caption{\label{fig:Fig4} Comparison between the evolution with $B_{||}$ of the calculated Fermi contour of electrons (a-d) and measured Fermi contour of $^4$CFs near $\nu=1/4$ (e-h). For simplicity, in (a-d) only the majority-spin contour is shown.  In (e-h), solid and dotted contours denote the $^4$CF Fermi contours for $\nu<1/4$ and $\nu>1/4$, respectively. Even though the electron Fermi sea completely splits at large $B_{||}$, the $^4$CF Fermi sea near $\nu=1/4$ remains intact.}
\end{figure}

The most surprising finding of our study emerges in Fig. 3 where we show the deduced $k_F^{*}$, normalized to $k_{F0}^{*}$, the value of $k_F^{*}$ at $B_{||}=0$. We observe a remarkable difference in the deduced $k_F^*$ for $\nu>1/4$ and $\nu<1/4$. For both cases, with increasing $B_{||}$, $k_F^{*}$ along $[\bar{1}10]$ increases while along $[110]$ it decreases. However, for $\nu<1/4$, the change is much slower compared to $\nu>1/4$. This is different from the $^2$CF Fermi sea at $\nu=1/2$ where both sides show similar anisotropy with increasing $B_{||}$ \cite{Kamburov.PRL.2013, Kamburov.PRB.2014}. The difference is particularly puzzling considering that $^2$CFs and $^4$CFs form in the same LL. 
 
Before discussing the asymmetry observed in Fig. 3 data, we emphasize that our measured Fermi sea anisotropy for $^4$CFs is qualitatively different from the \textit{electron} Fermi sea anisotropy at $B_{\perp} = 0$. The comparison is summarized in Fig. 4 where we show the Fermi contours of the electrons (top panels), calculated self-consistently based on the $8\times8$ Kane Hamiltonian \cite{Winkler.2003, Kamburov.PRB.2013, Mueed.PRL.Fermi2015}, and the $^4$CF Fermi contours deduced from our measurements (bottom panels). For electrons, the Fermi sea becomes severely distorted with increasing $B_{||}$ and even splits into two tear-shaped seas, signaling the formation of a bilayer system, as confirmed in experiments \cite{Kamburov.PRB.2013, Mueed.PRL.Fermi2015}. In stark contrast, the $^4$CFs Fermi sea is much less anisotropic and remains connected even at the highest $B_{||}=30$ T. This is similar to what was seen for $^2$CFs except that our measured Fermi sea anisotropy for $^4$CFs is even smaller than for $^2$CFs for the same quantum well width \cite {Kamburov.PRB.2014}. At $B_{||}=25$ T, e.g., $k_F^*/k_{F0}=1.9$ for the $^2$CFs \cite {Kamburov.PRB.2014} while the $^4$CFs exhibit $k_F^*/k_{F0} = 1.3$ and $1.6$ for $\nu<1/4$ and $\nu>1/4$, respectively.

For an understanding of the qualitative difference between the Fermi sea anisotropies of electrons and $^4$CFs, we use a simple model, inspired by Fermi liquid theory. Developed in Ref. \cite{Kamburov.PRB.2014} to explain $^2$CF data, this model takes into account the coupling of  $B_{||}$ to the out-of-plane (orbital) motion of the quasi-2D charged particles confined to a quantum well of width $w$, and provides an estimate for the Fermi sea anisotropy \cite{Stern.PRL.1968}. In the limit of a small anisotropy where this model is valid, it yields an elliptical Fermi contour with minor and major Fermi wave vectors:
\begin{equation} \label{Eq1}
k_{x,y} = \sqrt{\frac{n}{\pi}} \left(1- \frac{2^{10}}{3^5\pi^6}\frac{e^2B_{||}^2}{\hbar^2} \frac{w^4m_z}{m_{||}}\right)^{\pm1/4}.
\end{equation}
Here $B_{||}$ is along the $x$ direction, and $m_{||}$ and $m_z$ are the particles' effective mass in the 2D plane and out-of-plane, respectively. It is reasonable to expect that the physics of CFs characterizes the in-plane dynamics of the quasiparticles in our experiments. According to Fermi liquid theory, $m_{||}$ should then be approximately the effective mass of CFs that contains electron-electron interaction and is about unity \cite{Jain.2007, Halperin.PRB.1993, Du.PRL.1993, Manoharan.PRL.1994}. (All effective masses are in units of the free electron mass.) On the other hand, the quantized perpendicular motion of the quasiparticles giving rise to the formation of electric subbands should reflect the band dynamics which is characterized by the band mass of electrons in GaAs, $m_z = 0.067$. The approximate validity of this simple model for the $^2$CFs was demonstrated in Ref. \cite{Kamburov.PRB.2014} where the much smaller measured $^2$CF Fermi sea anisotropy compared to that of the zero-field electrons and its dependence on $w$ was explained.

In Fig. 3(b) we show the predictions of Eq. (1) (orange curves) using different values of $m_{||}$, with $m_z$ fixed at 0.067. The curve with $m_{||}=2.4$ fits the $\nu<1/4$ data reasonably well. For $\nu>1/4$, none of the curves fit the experimental data well \cite{footnote3}, but a comparison with the data suggests that $m_{||}$ is smaller than 2.4, namely that there is an \textit{asymmetry} between $m_{||}$ for $\nu<1/4$ and $\nu>1/4$. It is noteworthy that a qualitatively similar asymmetry in $^4$CF mass ($m^*$) was also deduced from measuring the temperature dependence of the strengths of fractional quantum Hall states near $\nu=1/4$ \cite{Pan.PRB.2000}. For $\nu>1/4$, the measured $m^*$ for $^4$CFs was found to be consistent with the value expected based on the $^2$CF mass (after scaling with $m^*\propto \sqrt{B_{\nu}}$ to take into account that $m^*$ is proportional to the Coulomb energy \cite{Jain.2007, Halperin.PRB.1993, Du.PRL.1993, Manoharan.PRL.1994}). However, for $\nu<1/4$, a much larger $m^*$ was deduced and was attributed to the formation of the pinned, magnetic-field-induced Wigner crystal (WC) which manifests as an insulating phase near $\nu=1/5$ \cite{Jiang.PRL.1990, Goldman.PRL.1990}.

For our sample, the expected $m^*$ for $^4$CFs near $\nu=1/4$ based on the $^2$CF $m^*$, and using the $m^* \propto\sqrt{B_{\nu}}$ scaling, is $\simeq$ 1.4  \cite{footnote2}. The data of Fig. 3(b) suggest that $m^*$ for $\nu<1/4$ is larger than this value, qualitatively consistent with the data of Ref. \cite{Pan.PRB.2000}. While the proximity to the WC formation as suggested in Ref. \cite{Pan.PRB.2000} and confirmed by numerical calculations \cite{Zu.PRB.1999} might be a possible explanation for the larger $^4$CF $m^*$ on the $\nu<1/4$ side in our sample also, we would like to emphasize an important point. The $m^*$ measured in Ref. \cite{Pan.PRB.2000} and calculated in Ref. \cite{Zu.PRB.1999} were in the range $0.237>\nu>0.222$, relatively close to the insulating phase that sets in at $\nu=0.21$ \cite{Jiang.PRL.1990, Goldman.PRL.1990, Pan.PRB.2000}. In contrast, we observe GR resistance minima in the range $0.246>\nu>0.242$, reasonably far from 0.21, and very close to $\nu=1/4$. It would therefore be surprising if the WC formation would affect the $^4$CFs so significantly \cite{SM}. We hope that our data would stimulate theoretical work for a quantitative understanding of $^4$CF properties and, in particular, the strong asymmetry we observe for the $^4$CFs Fermi sea anisotropy on the two sides of $\nu=1/4$.

Note added, recently we came to know of J. Wang's proposal of a flux-attached Dirac fermion theory where the Berry phase of composite fermion Fermi liquid at $\nu=1/4$ is computed \cite{Wang.PRL.2019}. This theory perhaps can provide an explanation of the asymmetry that we observe for the $^4$CFs.

\begin{acknowledgments}
We acknowledge support through the National Science Foundation (Grants DMR 1709076 and ECCS 1508925) for measurements, and the National Science Foundation (Grant No. MRSEC DMR 1420541), the U.S. Department of Energy Basic Energy Science (Grant No. DE-FG02-00-ER45841), and the Gordon and Betty Moore Foundation (Grant No. GBMF4420 for sample fabrication and characterization. This research is funded in part by QuantEmX travel grants from the Institute for Complex Adaptive Matter and the Gordon and Betty Moore Foundation through Grant No. GBMF5305 to M. S. H., M. K. M., and M. S. A portion of this work was performed at the National High Magnetic Field Laboratory, which is supported by National Science Foundation Cooperative Agreement No. DMR-1644779 and the State of Florida. We thank S. Hannahs, T. Murphy, J. Park, H. Baek, and G. Jones at NHMFL for technical support. We also thank J. K. Jain, W. Pan, M. Ippoliti, R. N. Bhatt, and J. Wang for illuminating discussions.
\end{acknowledgments}


\begin{thebibliography}{99}
\bibitem{footnote1} Throughout this Letter we use ``$*$'' to denote CF parameters.
\bibitem{footnote.n1} If the $^4$CFs were spin unpolarized, then $k_F^*$ would be $\sqrt{2\pi n}$, i.e., $\sqrt{2}$ times smaller than the fully-spin-polarized case, resulting in $\sqrt{2}$ times smaller $B_{i=1}^*$.
\bibitem{footnote.n2} Note that, in contrast to $\nu=1/2$ where the LL is less than half-filled on one side of $1/2$ and more than half-filled on the other side, the LL is less than half-filled on both sides of $\nu=1/4$, so the minority carriers are electrons on both sides and their density is equal to $n$.
\bibitem {footnote3} The experimental data also show a clear ``kink'' in the $^4$CF Fermi sea anisotropy near $B_{||}\simeq 25$ T; this is not explained by Eq. (1) either.
\bibitem {footnote2} Equation (1) also explains why $^4$CFs have a smaller Fermi sea anisotropy than $^2$CFs in the same sample at a given $B_{||}$. This is because $^4$CFs have a larger $m^*$.
\end{thebibliography}
\end{document}